\documentclass[aps,prb,twocolumn,superscriptaddress,showpacs]{revtex4}
\usepackage{graphicx}
\begin{document}
\title{
Ground state and edge excitations of quantum Hall liquid at
filling factor 2/3 }
\author{Zi-Xiang Hu}
\affiliation{Zhejiang Institute of Modern Physics, Zhejiang
University, Hangzhou 310027, P.R. China} \affiliation{National
High Magnetic Field Laboratory and Department of Physics, Florida
State University, Tallahassee, Florida 32306, USA}
\author{Hua Chen}
\affiliation{Zhejiang Institute of Modern Physics, Zhejiang
University, Hangzhou 310027, P.R. China}
\author{Kun Yang}
\affiliation{National High Magnetic Field Laboratory and
Department of Physics, Florida State University, Tallahassee,
Florida 32306, USA}
\author{E. H. Rezayi}
\affiliation{Department of Physics, California State University
Los Angeles, Los Angeles, California 90032, USA}
\author{Xin Wan}
\affiliation{Asia Pacific Center for Theoretical Physics and
Department of Physics, Pohang University of Science and
Technology, Pohang, Gyeongbuk 790-784, Korea}
\date{\today}

 \begin{abstract}

 We present a numerical study of fractional quantum Hall liquid at
 Landau level filling factor $\nu=2/3$ in a microscopic model
 including long-range Coulomb interaction and edge confining
 potential, based on the disc geometry. We find the ground state is
 accurately described by the particle-hole conjugate of a $\nu=1/3$
 Laughlin state. We also find there are two counter-propagating edge
 modes, and the velocity of the forward-propagating mode is larger
 than the backward-propagating mode. The velocities have opposite
 responses to the change of the background confinement potential. On
 the other hand changing the two-body Coulomb potential has
 qualitatively the same effect on the velocities; for example  we
 find increasing layer thickness (which softens of the Coulomb
 interaction) reduces both the forward mode and the backward mode velocities.

 \end{abstract}
  \maketitle

 \section{introduction}
 The fractional quantum Hall (FQH) effect is a remarkable phenomenon
 observed in two-dimensional electron gases (2DEGs) in a strong perpendicular
 magnetic field. FQH liquids are  gapped and believed to possess topological
 order.~\cite{wenbook} In particular, it has been established that the
 $\nu=1/3$ Laughlin state represents an Abelian topological phase. The
 excitations in such a phase can carry a fraction of electron charge
 and have fractional statistics which are in-between bosonic and fermionic
 statistics. Various experiments have reported observation of
 fractional
 charge~\cite{Science.267.1010,Nature.389.162,PhysRevLett.79.2526}.
 Recently, a series of experiments~\cite{camino:246802,camino:075342}
 observed the so-called superperiods in the conductance oscillations
 in an FQH quasiparticle interferometer, which have been interpreted
 as a reflection of fractional statistics.~\cite{goldman:045334,
 kim:216404} The bulk topological order is also reflected in the
 corresponding edge excitations, which are gapless.  For a $\nu=1/3$
 FQH liquid, with a sharp confining potential (no edge reconstruction), 
 there is only a single branch of bosonic excitations at the edge.
 The bosonic edge mode is chiral, i.e., propagating along the edge in
 one direction (determined by the ${\bf E}\times {\bf B}$ drift) only,
  because the magnetic field breaks the time-reversal symmetry.  The
 edge physics can be described by the chiral Luttinger liquid theory
 and has been verified by numerical tests in microscopic
 models.~\cite{palacios96,PhysRevB.68.125307}

\begin{figure}
 \includegraphics[width=7cm]{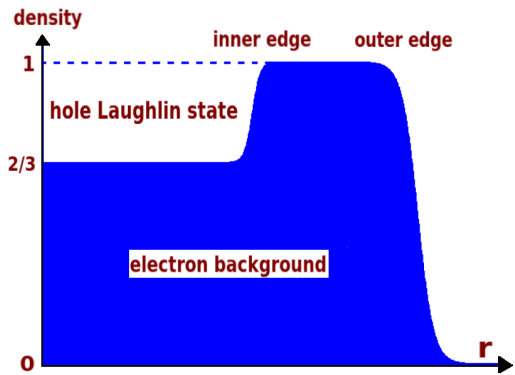}
 \caption{\label{edge}(color online) Schematic picture of
 electron density profile along radial direction of a $\nu=2/3$
 FQH droplet. We assume there is a hole Laughlin state embedded in
 the $\nu=1$ electron background. Therefore, there are two
 interfaces: one is between the 1/3 hole Laughlin state and the
 integral filling background and the other is between the $\nu=1$
 integer quantum Hall droplet and the vacuum.}
 \end{figure}

 In a hierarchical state, the FQH liquid supports multiple branches
 of edge excitations. Depending on the bulk topological order, the
 edge modes may propagate in the same direction or in opposite
 directions. The simplest case with counter-propagating edge modes
 is the spin-polarized FQH liquid at filling fraction $\nu=2/3$,
 which can be regarded as the particle-hole conjugate of a
 $\nu=1/3$ Laughlin state or, equivalently, a hole Laughlin state
 embedded in a $\nu=1$ integer quantum Hall background (see
 Fig.~\ref{edge} for an
 illustration).~\cite{PhysRevLett.64.220,PhysRevLett.67.2060,
 wen:IJMPB} Roughly speaking, the inner and outer edges are located
 at density changes of $2/3\rightarrow 1$ and $1\rightarrow 0$, respectively.
 In general, the two edge modes are coupled to each other and their
 properties may be dominated by disorder in the presence of random
 edge tunneling.~\cite{PhysRevB.51.13449,PhysRevLett.72.4129} The
 edge physics of the $\nu=2/3$ state is intriguing since one of the edge
 modes propagates opposite to the classical skipping orbits
 dictated by the uniform magnetic field, leading to a negative
 contribution to thermal Hall conductivity.~\cite{kane97} The
 counter-propagating edge modes have since been
 studied~\cite{PhysRevLett.74.2090,PhysRevB.52.17393,
 PhysRevLett.77.2538,PhysRevB5813778,JETPLetters82.539} both
 theoretically and experimentally in recent years.

 Recently, a similar but more delicate situation arises at filling
 fraction $\nu = 5/2$, where the Moore-Read Pfaffian
 state~\cite{moore91} and its particle-hole conjugated state,
 dubbed the anti-Pfaffian state,~\cite{lee:236807,levin:236806}
 compete for the ground state. In the absence of Landau level
 mixing, impurity, or edge confinement, the particle-hole
 symmetry is unbroken. In this case the two states, in the bulk, are
 expected to be degenerate in the thermodynamic limit. But
 these two states have very different edge structures: the Pfaffian
 state supports two co-propagating chiral edge modes (a charged
 bosonic mode and a neutral fermionic mode), while the anti-Pfaffian
 state supports three counter-propagating charge and neutral
 modes.~\cite{lee:236807,levin:236806} Their relation is somewhat
 similar to that between the $\nu=1/3$ and $\nu=2/3$ edge states. We
 note very recent experiments~\cite{dolev08,radu08,willett08} found
 indications of quasiparticle excitations with charge $e/4$ supported
 by both states and, interestingly, tunneling
 experiments~\cite{radu08} seem to favor the anti-Pfaffian state over
 the Pfaffian state. We also note that the particle-hole conjugates
 of the Read-Rezayi state~\cite{PhysRevB.59.8084} have been studied
 theoretically, with emphasis on properties of their edge
excitations.\cite{bishara:241306}

 Motivated by the recent work on the particle-hole conjugate of the
 Pfaffian state~\cite{lee:236807,levin:236806} and Read-Rezayi
 states,\cite{bishara:241306} as well as by the experimental
 measurement of the $I-V$ spectroscopy between individual edge
 channels,~\cite{arXiv:0803.2612v1} we revisit the polarized $\nu =
 2/3$ FQH state with a detailed numerical study on the edge modes
 of the $\nu=2/3$ FQH droplet using a semi-realistic microscopic
 model. We find the ground states of the system for a wide parameter range
 are well described by the composite of a $\nu = 1$ IQH droplet and
 a $\nu = 1/3$ Laughlin hole droplet. The number of electrons and
 holes in the two droplets vary as the strength of confining
 potential varies, under the constraint that the total number of
 electrons does not change.  Two counter-propagating edge modes are
 clearly visible in our results. Quantitatively, we find the
 forward-propagating outer edge mode (arising from the IQH edge)
 has a larger velocity than that of the backward-propagating inner
 edge (from the hole FQH edge). The structure of the excitation
 spectrum of the inner edge is identical to that of the Laughlin
 state at $\nu=1/3$ except for direction of propagation. Increasing
 the edge confining potential increases the outer edge mode
 velocity and {\em reduces} the inner edge mode velocity. We also
 carry out a particle-hole transformation of the electronic
 Hamiltonian with hard-core interaction to generate the Hamiltonian
 that makes the hole Laughlin state and hole edge states as its exact
 zero-energy ground states. Using a mixed Hamiltonian which
 contains both the electron Coulomb Hamiltonian and the the
 conjugate Hamiltonian of the two-body hard-core interaction, the
 bulk excitation energies can be raised to allow for a clearer
 separation between bulk and edge excitations. We find our
 results are robust in the presence of the electronic layer
 thickness, whose main effect is softening the Coulomb interaction
 and reducing the velocities of both edge modes.

 The rest of the paper is organized as follows. In
 Sec.~\ref{sec:model} we describe the microscopic model used in
 this work. We discuss the nature of the ground states in
 Sec.~\ref{sec:groundstates}. We present the overlap study of the
 ground states with variational wave functions in
 Sec.~\ref{sec:overlap}. A detailed analysis of the edge states
 follows in Sec.~\ref{sec:edgestates}. One can find explicit
 construction of the Hamiltonian for the variational wave functions
 of the ground states and edge states in
 Sec.~\ref{sec:phtransformation}. We consider the effect of electron
 layer thickness in Sec.~\ref{sec:layerthickness} and quasihole
 excitations in Sec.~\ref{sec:quasiholes}. Finally, some concluding
 remarks are offered in Sec.~\ref{sec:conclusion}.

 \section{The model}
 \label{sec:model} In recent years, we have developed a semi-realistic
 microscopic model for FQH liquids and have studied edge excitations
 and instabilities, quasihole/quasiparticle excitations, and edge
 tunneling in Laughlin and Moore-Read Pfaffian
 phases.~\cite{PhysRevLett.88.056802,PhysRevB.68.125307,wan05,
   wan:256804,wan:165316,hu:075331} The advantage of the model is
 that, depending on the parameters, the Laughlin phase, the Moore-Read
 phase, as well as edge reconstructed states and
 quasihole/quasiparticle states, emerge naturally as the global ground
 state of the microscopic Hamiltonian without any explicit
 assumptions, e.g., on the value of the ground state angular momentum.
 This way, we can study the stability of phases and their competition.
 Another advantage of the model is that we can analyze the edge
 excitations of the semi-realistic system and identify them in a
 one-to-one correspondence with edge excitations of the corresponding
 edge theory (or conformal field theory). In addition to confirming
 the bulk topological order, we can use the microscopic calculation to
 extract energetic quantities, such as edge velocities, which are
 crucial for quantitative comparisons with experiments; for example in
 a recent study we used the edge mode velocities extracted from our
 numerical study to estimate the quasiparticle dephasing length at
 finite temperatures at $\nu=5/2$.~\cite{wan:165316} In this paper, we
 apply our model and methods to the $\nu = 2/3$ FQH system.

 In this model, we consider a 2DEG confined to a plane with rotational
 symmetry. There is a neutralizing background charge distributed
 uniformly on a parallel disk of radius $R$ at a distance $d$ above
 the 2DEG (see Fig. 1 of Ref.~\onlinecite{PhysRevB.68.125307} for an
 illustration). The total charge of the disc is $N_e e$, where $N_e$
 is the number of electrons confined to the plane and the radius $R$
 or, equivalently, the density of the background charge is determined
 by the filling fraction $\nu$. We consider $\nu=N_e/N_\Phi=2N_e
 l_B^2/R^2$, where $l_B$ is the magnetic length and $N_\Phi$ is the
 number of flux quanta enclosed in the disc. The distance $d$
 parameterizes the strength of the electron confining potential due to
 attraction from background charge, which becomes weaker as $d$
 increases. We assume the electrons are spin-polarized, which is the
 case in strong magnetic fields. In the second quantization language,
 the Hamiltonian is written as: \begin{equation} \label{HC} H_C =
   \frac{{\rm{1}}}{{\rm{2}}}\sum\limits_{{\{m_i\}}} {V_{m_1m_2m_3m_4}
     c_{m_1}^ + c_{m_2}^ + c_{m_4} c_{m_3} } + \sum\limits_m {U_m c_m^
     + c_m }
 \end{equation}
 where the Coulomb matrix elements $ V_{\{m_i\}}$ are
 \begin{equation}
   V_{\{m_i\}}=\int d^2 r_1\int d^2 r_2 \phi _{m_1}^*
 (\vec{r}_1 )\phi_{m_2}^* (\vec{r}_2)
      \frac{{e^2 }}{{\varepsilon r_{12} }}
 \phi _{m_3} (\vec{r}_1)\phi _{m_4} (\vec{r}_2),
 \end{equation}
 and the background confining potential $U_m$ as a function of $d$
 is
 \begin{equation}
  U_m = {N_e e^2 \over \pi R^2 \varepsilon } \int d^2 r \int_{\rho
  < R} d^2 \rho \frac{|\phi _m (\vec{r})|^2}{\sqrt{|\vec{r} -
 \vec{\rho}|^2 + d^2}}. \end{equation} Here $\epsilon$ is the
 dielectric constant. We use the symmetric gauge $\vec{A}=(-
 \frac{By}{2},\frac{Bx}{2})$; the single-particle wave function
 $\phi_m$ in the lowest Landau level is: \begin{equation} \phi_m(z) =
 (2\pi2^m m!)^{-1/2} z^m e^{-|z|^2/4}. \end{equation} Throughout the
 paper, we use the magnetic length $l_B$ as length unit and
 $e^2/\epsilon l_B$ as energy unit.

 It is often convenient to cast the Coulomb matrix elements into a
 weighted sum of pseudopotentials introduced by
 Haldane.~\cite{haldane83} One of the advantages of expressing
 two-body interactions in terms of pseudopotentials is that the
 Laughlin states become exact ground states for specific
 pseudopotential Hamiltonians; for example at filling fraction $\nu =
 1/3$, with hard-core interaction between electrons (in Haldane's
 pseudopotential language, $V_m = \delta_{1,m}$) and in the absence of
 confining potential, the Laughlin state~\cite{PhysRevLett.50.1395}
\begin{equation}\label{laughlin}
 \Psi _{1/3} (z_1 , \cdots z_N ) = \prod\limits_{i > j}^N {(z_i  -
 z_j )^3 } \exp \{ - \frac{1}{4}\sum\limits_{i = 1}^N {|z_i |^2 }
 \}
 \end{equation}
 is the exact ground state with zero energy, which exists in the
 subspace of total angular momentum $M=3N_e(N_e-1)/2$, for $N_e$
 electrons in at least $N_{orb}=3N_e-2$ orbitals.

 \section{Ground State quantum numbers}
 \label{sec:groundstates}

 To begin our study on the $\nu=2/3$ system, we ask to what extent we
 can conclude that the ground state of the semi-realistic model can be
 described by the particle-hole conjugate of the 1/3 Laughlin state on
 an IQH background.  The schematic profile of the electron density is
 shown in Fig.~\ref{edge} in which  we
 neglect the density oscillation of the Laughlin state for holes near
its edge (see realistic curves in Fig.~\ref{density} for details).
Suppose
 the system contains $N_e$ electrons filling up to the $N_I =$th
 orbital
 (with single particle angular momentum $m = 0, 1, ..., N_I-1$).
 According to this picture, we have two droplets: $N_I$ electrons
 fill the lowest Landau level (LLL) and form a $\nu_I=1$ IQH state;
 in addition, $N_h = (N_{I} - N_e)$ holes form a $\nu_h=1/3$ hole
 Laughlin state.  The total angular momentum for such a state is
\begin{equation}\label{qm}
 M = \frac{N_I(N_I-1)}{2} -
 \frac{3(N_I-N_e)(N_I-N_e-1)}{2}.
 \end{equation}
 To reveal such a state in a microscopic calculation, one needs $N_{orb}
 \ge N_I$ orbitals.  For example, with $N_e = 20$ electrons filling
 $N_I = 26$ orbitals, we have a 6-hole Laughlin droplet (filling the
 innermost 16 orbitals) and a $\nu_I = 1$ IQH droplet (filling all 26
 orbitals).  The total angular momentum of the state is $M = 280$. We
 note that for the specific system the average filling fraction is somewhat
 different from $\nu = 2/3$, due to the higher density near the edge.

 If the hole-droplet picture is correct, one can possibly (but not
 necessarily for energetic reasons) find that the global ground state
 of the semi-realistic model has a total angular momentum $M = 280$,
  if 20 electrons are distributed in $N_{orb} \ge N_I = 26$ orbitals.

 \begin{figure}
   \includegraphics[width=7cm, height=5cm]{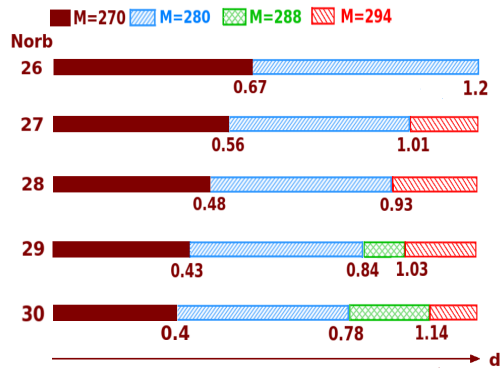}
 \caption{\label{phase}(color online) Phase diagram for systems with
 20 electrons at filling factor $\nu=2/3$, as a function of number of
 single electron orbitals $N_{orb}$ and background charge distance
 $d$. The ground state total angular momentum $M$ changes as d
 increases. Angular momenta $M=270, 280, 288, \mbox{and} 294$
 correspond to 5-, 6-, 7-, and 8-hole Laughlin ground states
 respectively. However the M=294 state with $N_{orb}=27$ is expected
 to be a stripe state which can be described by $|\Psi_{SP}\rangle =
|011111100000011111111111111\rangle$.}
 \end{figure}

 We study the global ground state of a system of 20 electrons with
 various $d$ and $N_{orb}$ (which serves as an additional hard edge
 confinement). We plot the results in Fig.~\ref{phase}. When the
 distance of the background charge $d$ increases, the confining
 potential becomes weaker and the total angular momentum of the
 global ground state increases and goes through steps at $M = 270$,
 280, 288, and 294. According to Eq.~(\ref{qm}), these states
 correspond to $N_h = 5$, 6, 7, and 8 holes, respectively. Unlike
 the Laughlin or Moore-Read Pfaffian cases where $M$ is uniquely
 determined by the number of electrons and a change of $M$ is an indication
 of instability,~\cite{PhysRevLett.88.056802,wan:256804} here we
 have a series of ``good" ground states, since the number of
 electrons only fixes the difference between the IQHE droplet and
 the hole droplet. In other words, the two edges can move
 simultaneously to respond to the change of edge confining
 potential. In this paper, we will focus on the 6-hole state,
 although the results should be general enough for other cases.

 The agreement between the actual series of ground state angular
 momenta and the prediction of Eq. (\ref{qm}) suggests that the
 picture of a Laughlin hole droplet embedded in an IQH droplet is a
 good description of the ground states at $\nu = 2/3$. In the
 following sections we will present more evidence that this is indeed
 true for most cases.  However, we would like to point out an
 exception here. The quantum number $M = 294$ is consistent with an
 8-hole state. But according to earlier discussion, we need $N_{orb}
 \ge 28$ orbitals to accommodate this state. The reader may have
 already noticed that it also appears in the $N_{orb} = 27$ case.
 This contradiction suggests that there is another state (or phase)
 that is competing with the hole-droplet picture. In fact, it even
 suppresses the 7-hole state ($M = 288$). Analysis of the electron
 occupation number suggests that this state is likely a stripe state.
 We can use a fermionic occupation number configuration to represent
 the state as $\vert \Psi_{SP} \rangle =
 |011111100000011111111111111\rangle$, where the string of 0s and 1s
 represents the occupation number of the corresponding single-particle
 orbital (from the left $m = 0$, 1, ..., $N_{orb}-1$). The stripe
 state is more compact than the 8-hole state and is expected to be
 energetically more favorable for the restricted case with $N_{orb} =
 27$ . To confirm this and to quantify the trend from $N_{orb}=27$ to
 30, we calculate the overlap between the ground state with $M = 294$
 and $\vert \Psi_{SP} \rangle$ and find that the ground state indeed
 has a large overlap
 ($|\langle\Psi_{SP}|\Psi_{M=294}\rangle|^2=25.4\%$) with a stripe
 state when $N_{orb}=27$ and $d=1.2l_B$. This overlap decreases as we
 increase the number of orbitals to $19.4\%$ in 28 orbitals, $16.2\%$
 in 29 orbitals, and $8.2\%$ in 30 orbitals, suggesting the ground
 state for large enough $N_{orb}$ is not the stripe phase, but
 possibly the 8-hole state. Note that for well-defined overlap, we
 need to add the proper number of 0s to the right of the occupation
 number configuration $\vert \Psi_{SP} \rangle$.

 \section{ground state wave functions and overlaps}
 \label{sec:overlap}

 In this section we study the ground state wave functions at
 $\nu = 2/3$ and show that they are consistent with the picture of a
 $\nu_h = 1/3$ hole droplet on top of a $\nu_e = 1$ electron
 droplet. For comparison between the two, we construct variational
 wave functions for $\nu = 2/3$ ground states by particle-hole
 conjugation of electron Laughlin states and calculate overlaps
 between them. To be specific, we target the 6-hole ground state with
 $M = 280$. We first fill the lowest 26 single-particle LLL orbitals
 to obtain an IQH state, which can be represented by a string of
 single-particle occupation numbers $\vert 111\cdots 111 \rangle$.
 Then, we construct a 6-hole Laughlin wave function in the following manner.
 A 6-electron Laughlin state, which is partially occupying the lowest
 16 orbitals, can be written as \begin{equation} \vert L^6_{16}
 \rangle = \prod_{1 \le i < j \le 6} (z_i - z_j)^3, \end{equation}
 where we have omitted a normalization constant and the Gaussian
 factor $\exp\{-\sum_i \vert z_i \vert^2/4\}$. We note that the Laughlin
 state is a many-body wave function and can be written,
 symbolically in the second quantization form, as
 \begin{equation}
 \vert L^6_{16} \rangle = \sum_{\{i_n\}}
 \alpha_{\{i_n\}}c^+_{i_1}c^+_{i_2}c^+_{i_3}c^+_{i_4}c^+_{i_5}c^+_{i_6} \vert
0 \rangle_{16}, \end{equation} where \begin{equation} \vert 0
\rangle_{16} = \vert 000 \cdots 000 \rangle_{16} \end{equation} is
the vacuum in 16 orbitals. Therefore, the 6-hole droplet embedded
in 16 orbitals is \begin{equation} \vert \bar{L}^6_{16} \rangle =
\sum_{\{i_n\}}
\alpha_{\{i_n\}}c_{i_1}c_{i_2}c_{i_3}c_{i_4}c_{i_5}c_{i_6} \vert
111 \cdots 111\rangle_{16}, \end{equation} which contains 10
electrons. After adding the additional 10 filled orbitals, we have
a many-body variational wave function for 20 electrons in 26
orbitals, which we denote as $|\bar{L}_{16}^6 \rangle^{20}_{26} =
\vert \bar{L}_{16}^6 \rangle \otimes \vert
1111111111\rangle_{10}$. For $N_{orb} > 26$, we can add trailing
0s for the empty orbitals at the edge accordingly, which we denote
as $|\bar{L}_{16}^6 \rangle^{20}_{N_{orb}}$. We compare
$|\bar{L}_{16}^6 \rangle^{20}_{N_{orb}}$ to the ground state of
our semi-realistic model with $d = 0.7 l_B$, at which all ground
states for $N_{orb}= 26$-30 have $M = 280$, as shown in
Fig.~\ref{phase}.  Table~\ref{table:overlap d0.7} shows the
overlap between the global ground state $|\Psi_{gs}\rangle$ in
different numbers of orbitals and the 6-hole variational wave
function $|\bar{L}_{16}^6 \rangle^{20}_{N_{orb}}$. As $N_{orb}$
varies from 26 to 30, the dimension of the Hilbert space increases
by a factor of about 450, while the overlap still survives at
about 60\%. The decrease is largely due to the fact that the outer
edge is no longer sharp as the angular momentum cut-off ($N_{orb}
- 1$) increases.

 \begin{table}
 \begin{center}
 \begin{tabular}{l|ccccc}
 $N_{orb}$ & \hspace{0.4cm}26\hspace{0.4cm} &
 \hspace{0.4cm}27\hspace{0.4cm} & \hspace{0.4cm}28\hspace{0.4cm} &
 \hspace{0.4cm}29\hspace{0.4cm} & \hspace{0.4cm}30\hspace{0.4cm}
 \\ \hline \hline
 \hspace{0.15cm}Size of HS\hspace{0.1cm} & 1123 & 10867 & 54799 &
 184717 & 473259
 \\
 $|\langle \Psi_{gs}|
 \bar{L}_{16}^6\rangle_{N_{orb}}^{20}|^2$\hspace{0.1cm} & 0.9401 &
 0.7868 & 0.7012 & 0.6431 & 0.6011
 \end{tabular}
 \end{center}
 \caption{ \label{table:overlap d0.7} Overlaps between 20-electron
 ground states with $M=280$ and $d=0.7l_B$ and the particle-hole
 conjugate of the 6-electron Laughlin state obtained from the hard-
 core Hamiltonian for several different numbers of orbitals
 $N_{orb}$. In the case of $d=0.7$, the ground state with  $M=280$ is
 the global ground state for all cases in Fig. \ref{phase}. The
 Hilbert subspace (HS) size for $M=280$ increases rapidly when
 $N_{orb}$ increases from 26 to 30 by about 450 times; however the
 overlap decreases slowly, indicating the robustness of the state.} \end{table}

\begin{figure}
  \includegraphics[width=8cm]{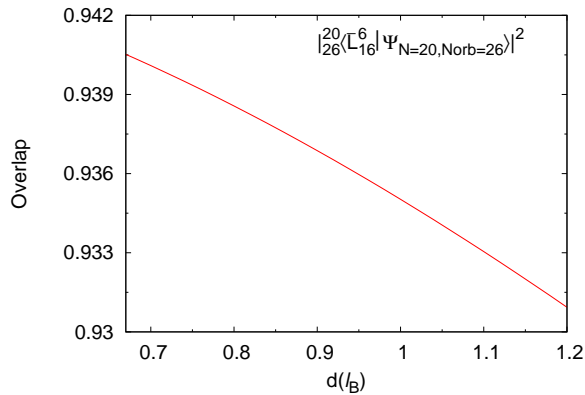}
 \caption{\label{evolution} Overlap between the
 20-electron ground state in 26 orbitals and the particle-hole
 conjugate of the 6-electron Laughlin state as a function of $d$.
 The decrease of the overlap indicates that the particle-hole
 conjugate Laughlin state favors smaller $d$, i.e., stronger confinement.}
 \end{figure}

 In addition, the strength of the confining potential due to
 neutralizing background charge also affects, though in a minor
 way, the overlap between the $\nu=2/3$ electron ground state and
 the variational wave function $|\bar{L}_{16}^6 \rangle^{20}_{N_{orb}}$.
 Fig.~\ref{evolution}
 shows that, for the regine in which the total angular momentum of
 the global ground state is $M=280$, the overlap for 20 electrons in
 26 orbitals decreases from
 0.94 to 0.93 as $d$ increases from 0.67 to 1.2. As the distance $d$
 between the 2DEG and the background charge increases, the confining
 potential the electrons experience becomes weaker. The electron wave
 function can expand, leading to a smaller overlap.  This is
 consistent with the $N_{orb}$ dependence, in the sense that the 
 hole-droplet ground state favors stronger confinement.  But since the
 range of $d$ for a ground state with a certain number of holes is
 narrow, we can neglect the 1\% change.

 \begin{figure}
  \includegraphics[width=8cm]{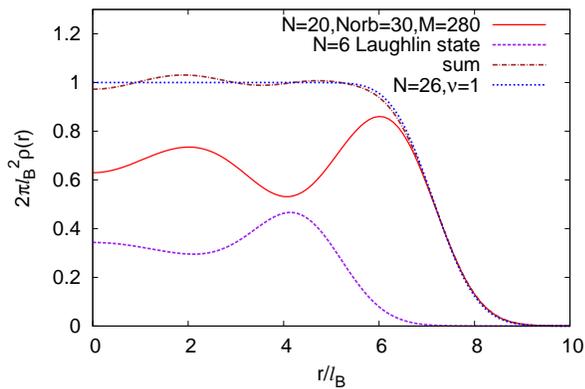}
 \caption{\label{density}(color online) Density profiles for the
 6-electron Laughlin state, the
 20-electron ground state in 30 orbitals with $M=280$ and $d=0.7l_B$,
 the sum of them, and the 26-electron IQH state. The sum
 is almost the same as the density of IQH.}
 \end{figure}

 An alternative way to compare states is to contrast the electron
 density profiles (though it is possible that two wave functions
 with identical density profiles can be orthogonal to each other).
 In Fig.~\ref{density}, we plot the density profile of a 6-electron
 Laughlin state and that of the 6-hole ground state obtained by
 exact diagonalization of a system of 20 electrons in 30 orbitals
 at $d = 0.7 l_B$. To compare, we plot the sum of the density
 profiles of the Laughlin state and the 6-hole state, together with
 the density profile of an IQH state with 26 electrons.  It is
 clear that the density sum is almost the same as the density of
 the IQH state, except for small oscillations which can be
 attributed to the long-range Coulomb interaction in the
 semi-realistic microscopic model.

 \section{edge excitations}
 \label{sec:edgestates} The analysis that can further confirm the
 picture that the $\nu = 2/3$ state is a Laughlin hole-droplet
 embedded in an IQH background is the study of edge excitations. In
 topological systems like FQH liquids, edge states have been
 demonstrated to be very effective and essential probes of the bulk
 topological order in both theoretical calculations and
 experiments. As shown in Fig.\ref{edge}, one expects  two
 counter-propagating edge modes originating from the two edges at
 $2/3\rightarrow 1$ and $1\rightarrow
 0$.~\cite{PhysRevLett.64.220,PhysRevLett.67.2060,wen:IJMPB} In the
 same spirit as the analysis in Refs.~\onlinecite{wan:256804} and
 \onlinecite{wan:165316}, we can label each low-energy edge
 excitation by two sets of occupation numbers $\{n_L(l_L)\}$ and
 $\{n_R(l_R)\}$ for the inner and outer edge modes with angular
 momenta $l_L$, $l_R$ and energies $\epsilon_L$, $\epsilon_R$,
 respectively. $n_L(l_L)$ and $n_R(l_R)$ are non-negative integers.
 The angular momentum and energy of an edge excitation, measured
 relatively from those of the ground state, are
 \begin{eqnarray}
 \Delta M &=& -\sum_{l_L} n_L(l_L)l_L + \sum_{l_R} n_R(l_R)l_R,
 \label{combination}\\
 \Delta E &=& \sum_{l_L} n_L(l_L)\epsilon_L(l_L) + \sum_{l_R}
 n_R(l_R)\epsilon_R(l_R).
 \label{combination1}
 \end{eqnarray}
 For the latter we assumed absence of interactions among the excitations.
 The negative sign in Eq.~(\ref{combination}) indicates the inner
 edge mode is propagating in the opposite direction to the outer edge
 mode.

\begin{figure}
  \includegraphics[width=8cm]{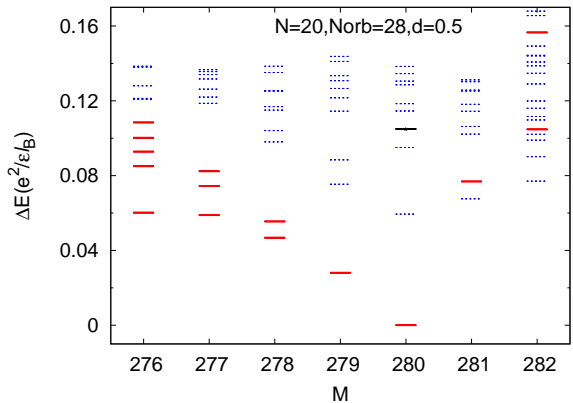}
 \caption{\label{twoedge}(color online) Low-energy spectrum for 20
 electrons in 28 orbitals with $d=0.5$. The inner and outer edge
 states are labeled by red bars and the simplest combination of
 the two is labeled by a black bar. Its energy (0.1049) is roughly
 the sum of the two lowest excitation energies for the two modes:
 0.02807(inner)+0.07687(outer). This state has a moderately large
 overlap(0.3959) with the particle-hole conjugate of a 6-electron
 Laughlin state with $\Delta M =1$ edge excitation and embedded in
 the IQH edge state with $\Delta M=1$: $|\bar{L}_{17}^6(\Delta
 M=1)\cdots 1101>$.}
 \end{figure}

 In Fig.~\ref{twoedge}, we plot the low-energy spectrum for 20
 electrons in 28 orbitals at $d=0.5l_B$, whose global ground state
 has $M = 280$. The edge states are labeled by red bars based on
 the analysis we will discuss in the following paragraphs.  Here we
 first point out that the inner edge excitations have negative
 $\Delta M$ and are separated by an energy gap from other states
 (presumably bulk states) in each angular momentum subspace. The
 number of these inner edge states (including the ground state) are
 1, 1, 2, 3, and 5 for $\Delta M=0$-4, as predicted by the chiral
 boson edge theory.~\cite{wen:IJMPB} They have significant overlap
 with a 6-hole Laughlin droplet with corresponding edge excitations
 embedded in a 26-electron IQH background. On the other hand, the
 outer edge excitations ($\Delta M > 0$) have higher excitation
 energies and are mixed with bulk states. In particular, for
 $\Delta M = 1$, the edge state is the second lowest eigenstate in
 the $M=281$ subspace. The state has a large overlap (62.7\%) with
 the 6-hole Laughlin droplet embedded in the 26-electron IQH state
 with an edge excitation at $\Delta M = 1$. Obviously, the outer
 edge mode has a larger velocity than the inner edge mode,
 consistent with the different charge density associated with the
 edge modes.

\begin{figure}
 \centering
 \includegraphics[width=8cm]{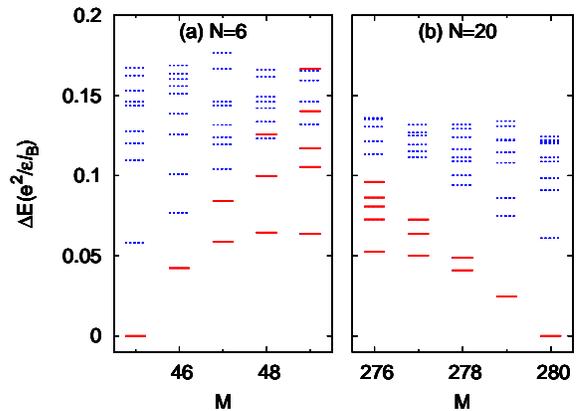}
 \caption{\label{compare}(color online) Comparison of the low energy
 excitation spectrum of the $\nu=1/3$, 6-electron Laughlin edge states
 (6 electrons in 22 orbitals with Coulomb interaction, $d =
 0.5l_B$) (a) and $\nu=2/3$, 6-hole Laughlin edge states(20
 electrons in 28 orbitals, $d=0.5l_B$) (b). The edge states are
 labeled by red bars. The overlaps between these edge states are
 shown in table \ref{table:compare}.}
 \end{figure}

 To identify the inner edge excitations, we compare the edge
 spectrum to that of a 6-electron system at $\nu=1/3$, with the
 neutralizing background charge at the same distance from the 2DEG.
 In Fig.~\ref{compare}, we plot side-by-side the $\nu=1/3$ and the
 $\nu=2/3$ edge spectra, both with $d=0.5l_B$. The edge states are
 labeled by red bars. From the comparison, one clearly sees that the
 $\nu = 2/3$ state and the $\nu = 1/3$ state have similar edge
 excitations, but along opposite directions. The one-to-one
 correspondence of the edge excitations in the two cases can be
 established by studying overlaps of the corresponding pairs. Of
 course, the overlap that we really calculated is the overlap between
 the eigenstate for $\nu = 2/3$ and the corresponding particle-hole
 conjugated state for $\nu = 1/3$ embedded in the 26-electron IQH
 background. To minimize the influence from the momentum cut-off, we
 choose 6 electrons in 22 orbitals with the same background
 ($d=0.5l_B$) for 1/3 filling. The results of the overlap calculation
 are summarized in Table~\ref{table:compare}. The overlap becomes
 smaller as we go to higher energy, but remains above 40\% up to
 $\vert \Delta M \vert = 4$. Overlaps between other pairs of states
 are significantly smaller.

 \begin{table*}
 \begin{center}
 \begin{tabular}{l|ccccc}
   & \hspace{0.5cm}1\hspace{0.5cm} & \hspace{0.5cm}2\hspace{0.5cm}
& \hspace{0.5cm}3\hspace{0.5cm} & \hspace{0.5cm}4\hspace{0.5cm} &
> \hspace{0.5cm}5\hspace{0.5cm}
 \\ \hline \hline
 $|\langle\Psi_{M=45}|\Psi_{M=280}\rangle|^2$\hspace{0.1cm} &
 $0.7527_{\langle1|1\rangle}$ &  & &  &
 \\
 $|\langle\Psi_{M=46}|\Psi_{M=279}\rangle|^2$\hspace{0.1cm} &
 $0.6615_{\langle1|1\rangle}$ &  & &  &
 \\
 $|\langle\Psi_{M=47}|\Psi_{M=278}\rangle|^2$\hspace{0.1cm} &
 $0.7003_{\langle1|1\rangle}$ & $0.5814_{\langle2|2\rangle}$ & &
 &
\\
 $|\langle\Psi_{M=48}|\Psi_{M=277}\rangle|^2$\hspace{0.1cm} &
 $0.7098_{\langle1|1\rangle}$ & $0.6147_{\langle2|2\rangle}$ &
 $0.5083_{\langle3|3\rangle}$ &  &
 \\
 $|\langle\Psi_{M=49}|\Psi_{M=276}\rangle|^2$\hspace{0.1cm} &
$0.7180_{\langle1|1\rangle}$ & $0.6238_{\langle2|2\rangle}$ &
 $0.6000_{\langle3|3\rangle}$ & $0.4999_{\langle5|4\rangle}$ &
 $0.4130_{\langle10|5\rangle}$
 \end{tabular}
 \end{center}
 \caption{ \label{table:compare}
  The overlaps between the particle-hole conjugated 6-electron edge
 states (in 22 orbitals with $d = 0.5l_B$; Fig.~\ref{compare}(a)) and
 the 20-electron inner edge states (in 28 orbitals with $d=0.5l_B$;
 Fig.~\ref{compare}(b)). The subscript $\langle n|m\rangle$ means
 this is the overlap between the $n'$th state in the former subspace
 and the $m'$th state in the later subspace. } \end{table*}

 In order to identify edge excitations with positive $\Delta M$, we
 need to consider the IQH edge excitations of the outer edge.  A
 26-electron IQH ground state in 28 orbitals can be represented by
 occupation numbers $\vert 11\cdots11100 \rangle$, with 26
 consecutive 1s followed by two 0s. So after adding a 6-hole
 Laughlin hole droplet, we denote the ground state as
$|\bar{L}_{16}^6 \cdots 11100\rangle$ for convenience, although
 the many-body variational state cannot be written as a single
 occupation number string (i.e., a Slater determinant). In the same
 spirit, we can construct and denote variational wave functions
 with IQH edge excitations as $|\bar{L}_{16}^6 \cdots 11010\rangle$
 for $\Delta M = 1$, $|\bar{L}_{16}^6 \cdots 11001\rangle$ and
 $|\bar{L}_{16}^6 \cdots 10110\rangle$ for $\Delta M = 2$ in order to
 emphasize the excitation at the outer edge.  We calculate the
 overlap between the variational wave function
 $|\bar{L}_{16}^6\cdots11010\rangle$ and the states in $M=281$
 subspace. The largest we find is the second state,
 $|\langle\Psi_{{M=281}_2}|\bar{L}_{16}^6...11010\rangle|^2=0.6272$,
 which we identify as the outer edge excitation. Similarly for
 $\Delta M = 2$, we identify the fifth state, which has
 $|\langle\Psi_{{M=282}_5}|\bar{L}_{16}^6...10110\rangle|^2=0.1896,$
 and the seventeenth state, which has
 $|\langle\Psi_{{M=282}_{17}}|\bar{L}_{16}^6...11001\rangle|^2=0.1794,
 $ as edge excitations. We note the overlap already becomes small for
 $\Delta M = 2$, indicating significant mixing between the edge
 states and bulk states. This is not surprising since, for the small
 system we consider, there is no gap protecting the edge states.

 Eqs.~(\ref{combination}) and (\ref{combination1}) suggest that
 there are also composite excitations that are combinations of
 these two counter-propagating edge modes. The simplest one is the
 combination of the edge states with $\Delta M=-1$ and $\Delta M=1$,
 which resides in the $M=280$ subspace. Intuitively, we can construct
 a variational wave function by particle-hole conjugating a
 6-electron Laughlin state with the $\Delta M = 1$ edge excitation
 and embedding it in the IQH state with $\Delta M = 1$, which we
 denote as $|\bar{L}_{17}^6 (\Delta M = 1)...11010\rangle$. We find
 that the fourth state in the $M=280$ subspace has the
 largest overlap (about 0.3959) with $|\bar{L}_{17}^6 (\Delta M =
 1)...11010\rangle$; meanwhile, its excitation energy ($\Delta E =
 0.1049$) is roughly the sum of the excitation energy of the
 $\Delta M=1$ edge state ($\Delta E = 0.07687$) for the outer edge
 and the $\Delta M=-1$ state ($\Delta E = 0.02807$) for the inner
 edge, confirming Eq.~(\ref{combination1}).

 \begin{figure}
 \centering
 \includegraphics[width=7cm]{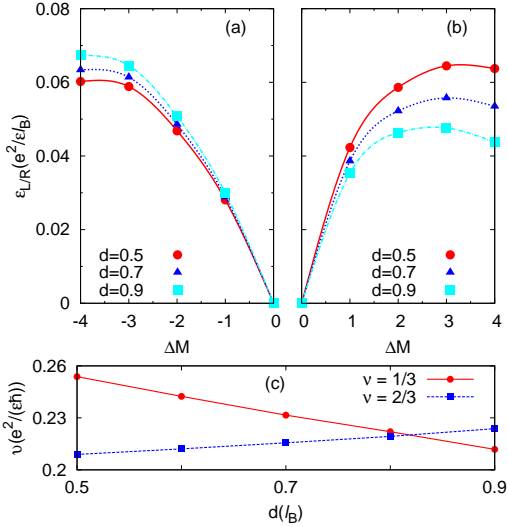}
 \caption{\label{velocityd}(color online) The dispersion relation
of the inner edge mode for 20 electrons in 28 orbitals at
 $\nu=2/3$ (a) and the edge mode of a 6-electron Laughlin
 droplet in 20 orbitals at $\nu=1/3$ (b) with different background
 confinement potentials. The evolutions of the edge velocities as a function
 of $d$ are plotted in (c). It can be seen that the velocity of
 the counter-propagating edge mode in the hole Laughlin state crosses
 that of the electron Laughlin state at around $d=0.82l_B$ and the
 velocity of the electron Laughlin state has an opposite response
 to the change of the background confinement to the hole Laughlin
 state. }
 \end{figure}

 Fig.~\ref{velocityd} compares the dispersion curves and
 corresponding edge velocities for both the inner edge mode of the
 20-electron droplet at $\nu = 2/3$ and the edge mode of a
 6-electron Laughlin droplet at $\nu = 1/3$ with different
 background confinement.  The velocity of an edge mode is defined
 as $v = |d\epsilon(k)/dk|$. The edge excitation with angular
 momentum $\Delta M$ measured from the ground state is related to
 the edge linear momentum $k=\Delta M/R$, where $R=\sqrt{3N_e}l_B$
 is the radius of the $N$-electron FQH droplet at $\nu = 2/3$ and
 $R=\sqrt{6N_e}l_B$ for $\nu = 1/3$. Here we smear the difference
 of the radius between the inner and outer edge of $\nu=2/3$.
 We find the velocity of the electron liquid $v_e$ is larger than
 the corresponding velocity $v_h$ for the hole droplet of the $\nu
 = 2/3$ FQH state at small $d$ (strong confinement), while $v_e$ is
 smaller than $v_h$ at large $d$ (weak confinement). The crossing
 happens at around $d = 0.82l_B$. At $d=0.5l_B$, the velocity of the
 electron edge
is
 about $0.25 e^2/(\epsilon \hbar)$, while
 the velocity of the hole edge mode is about $0.22 e^2/(\epsilon
 \hbar)$. As $d$ increases, $v_e$ decreases (roughly linearly)
 because the edge confinement is weaker and electrons tend to move
 out. However, $v_h$ increases linearly with $d$. Therefore, the
 symmetry correspondence between the electron droplet and the hole
 droplet is not exact in the presence of edge confinement.

\begin{figure}
  \includegraphics[width=8cm]{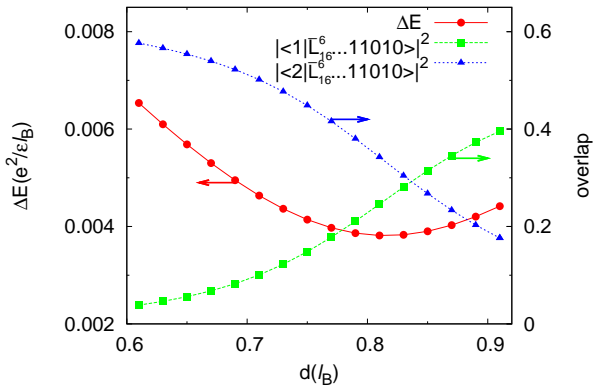}
 \caption{\label{fig8}(color online) The energy gap (red circle
 point line) $\Delta E=E_2-E_1$ and the overlap between the
 integral edge state $|\bar{L}_{16}^6...11010\rangle$ and the first
 two lowest states (square and triangular point line) in M=281
subspace as a function of d. The energy gap reaches its minima and
the overlap has a crossover between the the first state and
 the second state at around $d_c=0.84l_B$ indicating that the first state
 becomes the outer edge state after $d>d_c$.}
 \end{figure}

 Meanwhile, in Fig.~\ref{twoedge}, we pointed out that the second
 lowest eigenstate in the $M = 281$ momentum subspace is the outer
 edge state at $d=0.5l_B$. We can expect that the excitation energy
 of the edge state decreases with $d$ and may drop below that of
 the lowest energy state (presumably a bulk state). This is indeed
 the case, as illustrated in Fig.~\ref{fig8}.  When increasing the
 distance $d$ to the neutralizing charge background, we find an
 anticrossing behavior of the lowest two eigenstates, as their
 energy difference $\Delta E=E_2-E_1$ shows a minimum at the
 crossing scale $d_c=0.84l_B$. Overlap calculations reveal that
 below this $d_c$, the lowest-energy state has a smaller overlap
 with the corresponding variational wave function of the edge state
 ($|\bar{L}_{16}^6...11010\rangle$) than the second lowest state.
 On the contrary, above $d_c$, the lowest-energy state has a larger
 overlap with the variational edge state, which can be regarded as
 the ground state with an $\Delta M = 1$ outer-edge IQH excitation.

\begin{figure}
  \includegraphics[width=7cm]{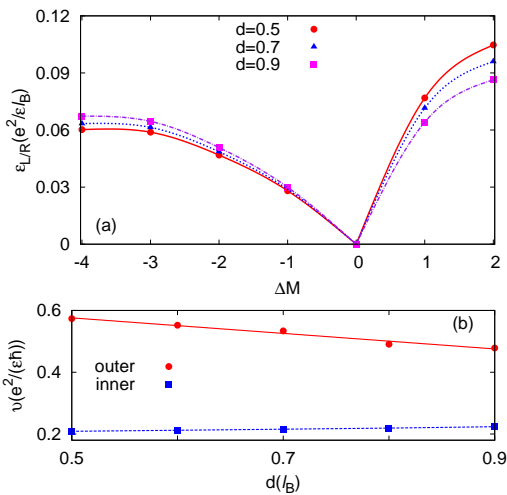}
 \caption{\label{out_velocity}(color online) The dispersion curves
 of a $\nu=2/3$ FQH droplet with 20 electrons in 28 orbitals (a)
 for both the inner and outer edge modes at different d's. As in
 the 6 electron Laughlin state, the velocities of the outer edge
 mode decrease linearly as a function of $d$ and  have larger
 velocities than the inner edge (b) in the whole parameter range
 where the global ground state resides in the M=280 subspace.}
 \end{figure}

 Fig.~\ref{out_velocity}(a) shows the edge dispersion curves for
 both the inner and the outer edge modes for different background
 confinement. Similar to the edge mode of the electron Laughlin
 state, the velocity of the outer edge mode of the $\nu=2/3$ FQH
 droplet decreases linearly as $d$ increases, as plotted in
 Fig.~\ref{out_velocity}(b). At $d=0.5l_B$, the velocity of the
 outer edge mode is about $0.596 e^2/(\epsilon \hbar)$,
 slightly smaller than 3 times the inner edge velocity.
 In this case the small
 deviation from 3 suggests the two edges may be weakly
 coupled.

 We conclude the section by pointing out that we can identify two edge
 modes for a $\nu = 2/3$ FQH droplet, propagating along opposite
 directions. The outer edge velocity is larger than the inner edge
 velocity. The outer and inner edge modes originate from electron and
 hole droplets, respectively, and show opposite dependence on the
 strength of the edge confining potential, which breaks the
 particle-hole symmetry.

 \section{particle-hole transformation}
 \label{sec:phtransformation}

 The Laughlin state is the exact
 zero-energy state of a special two-body Hamiltonian with hard-core
 interaction. In this section we use particle-hole transformation to
 construct Hamiltonians that make the hole Laughlin states (which we
 used as variational ground states in previous sections) exact ground
 states. As we are going to show below, such Hamiltonians include not
 only the same hard-core interaction but also an additional one-body
 term in the electron basis.

 We start by considering a generic two-body Hamiltonian in terms of
 {\em hole} operators: \begin{equation} H_h = {1\over
 2}\sum_{\{m_i=0\}}^{N_{orb}-
 1}V_{m_1m_2m_3m_4}h_{m_1}^+h_{m_2}^+h_{m_4}h_{m_3}, \end{equation}
 where the hole operators $h^+$ and $h$ are related to electron
 operators: $h^+ = c$ and $h = c^+$. It is straightforward to express
 the same Hamiltonian in terms of electron operators:
\begin{eqnarray}\label{holeHam}
   H_{h }
   &=&{1\over 2} \sum_{\{m_i=0\}}^{N_{orb}-1}V_{m_1m_2m_3m_4}c_{m_4}^+c_{m_3}^+c_{m_1}c_{m_2} \nonumber\\
   &-&\sum_m
   \bar{U}_m c_m^+ c_m+ {\rm const.},
\end{eqnarray}
 where
 \begin{equation}
 \bar{U}_m = \sum_{n}\{ V_{nmnm} -V_{nmmn}\}
 \end{equation}
 is the Hartree-Fock self-energy of the state $m$ when all the $N_{orb}$
 orbitals are occupied by electrons. Thus, in the electron basis we
 get the same two-body interaction {\em plus} a one body potential, which
 is {\em attractive} if the two-body potential is repulsive.

\begin{figure}
  \includegraphics[width=7cm]{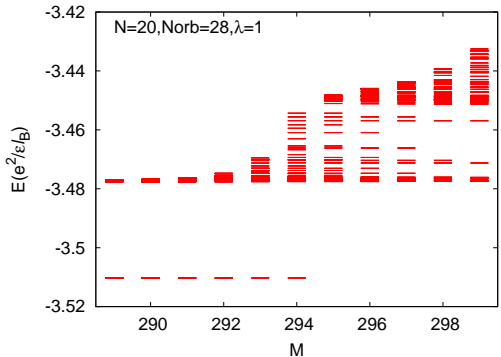}
 \caption{\label{phexact}(color online) The energy spectrum of the P-H
 conjugated Hamiltonian with hard-core interaction
 (Eq.~(\ref{holeHam})) for 20 electrons in the 28 orbitals. $M$ is
 the total angular momentum for the 20 electrons. The first 50 energy
 levels are plotted for each M. The lowest energy states (they should
 be the exact zero energy state if we include the constant term) at
 M=294,...,289 are degenerate with degeneracy 1,1,2,3,5,7.} \end{figure}

From now on we focus on the specific short-range (hard-core)
 interaction that corresponds to Haldane pseudopotential
 $V_m=\delta_{1,m}$ for $H_h$. After diagonalizing the Hamiltonian
 $H_h$ with $N = 20$ electrons in $N_{orb}=28$, we obtain the
 energy spectrum in Fig.~\ref{phexact}. It is worth pointing out that
 since there is no edge confinement other than the momentum cut-off
 due to the choice of $N_{orb}$, we have $N_h = N_{orb} - N = 8$
 holes in the system. Not surprisingly, the largest angular momentum
 of the degenerate ground states is $M_0 = 294$, as expected from
 Eq.~(\ref{qm}), with $N_I = N_{orb}=28$ and $N_e = 20$. This is the
 densest ground state configuration for holes and is thus
 incompressible. For $\Delta M = M - M_0 =$ -1, -2, ..., -5, we find
 the ground state degeneracy in each subspace is $n (\Delta M) = 1$,
  2, 3, 5, and 7, respectively. This is precisely the number expected
 for the Laughlin droplet (except for this case, the momentum is
 negative), as generated by \begin{equation} \sum_{\Delta M} n(\Delta
 M) q^{\Delta M} = \prod_{m = 1}^{\infty} {1 \over {1 - q^{-m}}}. \end{equation}

\begin{figure}
  \includegraphics[width=7cm]{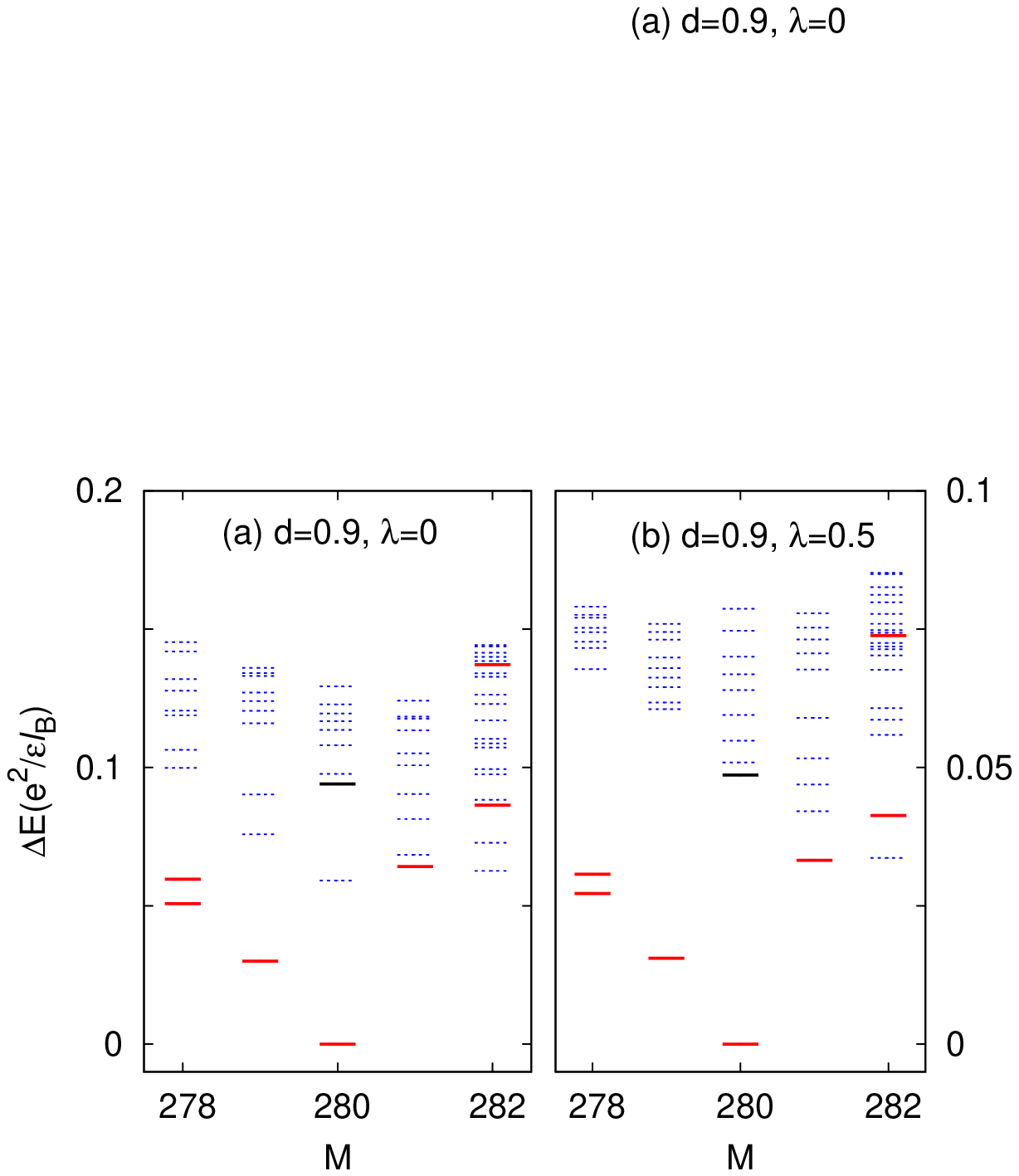}
  \caption{\label{mix}(color online) The energy spectrum for the pure
 Coulomb Hamiltonian with $\lambda =0$ (a) and a mixed Hamiltonian
 with  $\lambda = 0.5$ (b). When $\lambda =0$ the overlaps between
 the outer edge states (and ground state) and their corresponding
 conjugated states are:
  $|\langle\Psi_{{M=280}_1}|\bar{L}_{16}^6\rangle_{26}^{20}|^2 =
 0.684$,  $|\langle\Psi_{{M=281}_1}|\bar{L}_{16}^6...11010\rangle|^2=
 0.385$,  $|\langle\Psi_{{M=282}_3}|\bar{L}_{16}^6...10110\rangle|^2
 = 0.133$,
 $|\langle\Psi_{{M=282}_{15}}|\bar{L}_{16}^6...11001\rangle|^2 =
   0.158$;  the overlap for the simplest linear combination state is:
  $|\langle\Psi_{{M=280}_3}|\bar{L}_{17}^6(\Delta M=1)
 ...11010\rangle|^2 = 0.194$.  In the case of a mixed Hamiltonian
 with $\lambda =0.5$, they are:
  $|\langle\Psi_{{M=280}_1}|\bar{L}_{16}^6\rangle_{26}^{20}|^2 =
 0.66$,  $|\langle\Psi_{{M=281}_1}|\bar{L}_{16}^6...11010\rangle|^2 =
 0.539$,  $|\langle\Psi_{{M=282}_2}|\bar{L}_{16}^6...10110\rangle|^2
 = 0.218$,
 $|\langle\Psi_{{M=282}_{10}}|\bar{L}_{16}^6...11001\rangle|^2 =
   0.107$,  $|\langle\Psi_{{M=280}_2}|\bar{L}_{17}^6(\Delta M=1)
 ...11010\rangle|^2 = 0.296$.} \end{figure}

In the spirit of Refs.~\onlinecite{wan:256804} and
\onlinecite{wan:165316}, a mixed Hamiltonian that contains both the
Coulomb Hamiltonian $H_C$ used in earlier sections and the above
hard-core Hamiltonian $H_h$ parameterized by $\lambda$ is considered:
 \begin{equation}\label{mixH} H = \lambda H_{h} + (1-\lambda) H_C. \end{equation}
 The idea here is that the hard-core Hamiltonian $H_h$ raises the
 energy of bulk excitations while having little effect on the edges
 states, thus its presence helps separate the two energetically.
 Fig.~\ref{mix} shows the energy spectrum for 20 electrons in 28
 orbitals in the pure Coulomb case ($\lambda = 0$) and a mixed case
 ($\lambda = 0.5$). Although the ground state of $H_h$ for 20
 electrons in 28 orbitals is the 8-hole Laughlin state, for the
 mixed Hamiltonian above we obtain a different ground state with
 the same quantum number as the 6-hole Laughlin state ($M=280$).
 This is because the neutralizing background charge at $d=0.9$ serves
 as a repulsive potential to the holes that pushes two holes to the
 outer edge. It is clear that edge excitations show up at lower
 energies for the case $\lambda = 0.5$. In particular, overlap
 calculations indicate that in the case of pure Coulomb interaction
 with $\lambda = 0$, the third state in M=280 subspace is the
 simplest combination state which has the largest overlap (0.194)
 with $|\bar{L}_{17}^6(\Delta M=1)...11010\rangle$, while in the case
 of the mixed Hamiltonian with $\lambda = 0.5$, it is the second
 state in M=280 subspace that has the largest overlap (0.296) with
$|\bar{L}_{17}^6(\Delta
 M=1)...11010\rangle$. Therefore, the mixing of the hard-core
 Hamiltonian indeed has the effect of separating the band of edge
 states from the
bulk states.

 \section{layer thickness}
 \label{sec:layerthickness}

 One improvement in the semi-realistic model is to consider the
 effect of finite electron layer thickness. In real experimental
 samples, quasi-two-dimensional electrons are confined in the GaAs
 quantum well, whose width can be as large as 30 nm or a few
 magnetic lengths. Since the vertical motion
 of electrons is suppressed at low temperatures, the quasi-2DEG can
 be approximated, to the lowest order, by an ideal 2DEG located at
 the peak of the wave function in the perpendicular direction. The
 finite width softens the repulsion between electrons and thus,
 together with other factors
 (like higher Landau level), may help stabilize certain fragile FQH
 states.~\cite{peterson08a,arXiv:0801.4819} In this section, we
briefly discuss the effects of the 2DEG layer thickness on the
 velocities of $\nu=2/3$ edge modes.  We  use the Fang-
 Howard~\cite{RevModPhys.54.437,PhysRevB.30.840} variational wave
 function, \begin{equation}  Z_0(z)=2(2b)^{-3/2}ze^{-z/2b}, \end{equation}
 to model the electron layer thickness effect, where $b$ is a measure
 of the well thickness.  We obtain the same qualitative behavior for
 an infinite quantum well
potential.

\begin{figure}
 \includegraphics[width=6cm]{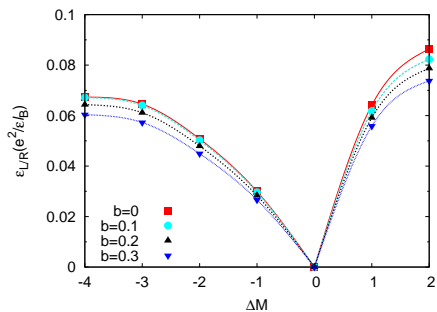}
 \caption{\label{thickness_v}(color online) The dispersion curves
 for both the inner edge and the outer edge mode for 20 electrons
 in 28 orbitals with $d=0.9l_B$ and different layer thicknesses.
 The the layer thickness softens the interaction between the
 electrons and reduces the edge mode velocities.}
 \end{figure}

 We can integrate the Fang-Howard wave function to obtain the
 renormalized Coulomb interaction in Fourier space \begin{equation}
 v_{FH}(k)=\frac{e^2}{\epsilon}\frac{1}{8k}\frac{3(kb)
 ^2+9kb+8}{(kb+1)^3}. \end{equation} In Fig.~\ref{thickness_v}, we
 show the edge dispersion curves for both the inner and the outer
 edge modes for a $\nu=2/3$ FQH droplet of 20 electrons with
 different layer thicknesses for $d =
 0.9$. The ground state angular momentum is again $M = 280$,
  consistent with the 6-hole Laughlin droplet picture. We find that
the velocities of both edge modes are reduced by increasing layer
 thickness, as expected. This is in contrast to the effects of the
 confining potential, which modifies edge velocities in an opposite way.

 \section{quasiparticle}
 \label{sec:quasiholes}

 In this section, we demonstrate that both one quasihole and one
 quasiparticle can be excited at the center of the FQH droplet with
 an additional short-range impurity potential. This section is a
 natural generalization of similar works by some of the authors for
 the Laughlin case at $\nu = 1/3$~\cite{hu:075331} and for the
 Moore-Read Pfaffian case.~\cite{wan:256804,wan:165316}

\begin{figure}
  \includegraphics[width=7cm]{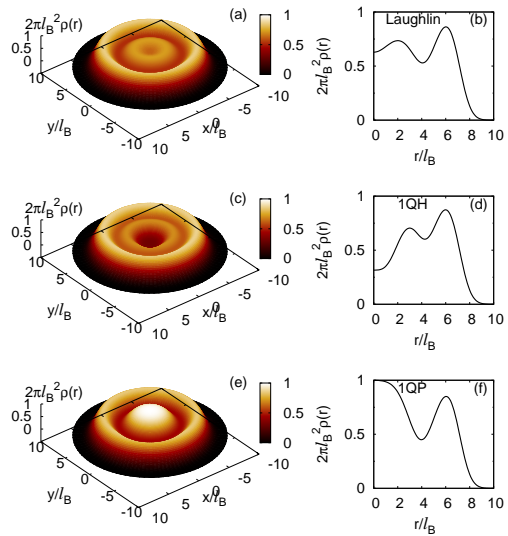}
 \caption{\label{quasi}(color online) The density profile of the
 2/3 state ( 1/3 hole-Laughlin state) and its quasihole and
 quasiparticle excitations. The system contains 20 electrons in 30
 orbitals with $d=0.5l_B$ and the Gaussian tip potential
 $H_W=W_g\sum_m\exp(-m^2/2s^2)c_m^+c_m$ has a width $s=2l_B$. (a) and
 (b) are the density profiles for the ground state with angular
 momentum M=280, (c) and (d) are the density profiles for one quasihole state with M=286 when $W_g=0.2$ and (e) and (f) are for one
 quasiparticle state with M=274 and $W_g=-0.2$.} \end{figure}

 We follow the previous work~\cite{hu:075331} by using a Gaussian
 impurity potential $H_W=W_g\sum_m\exp(-m^2/2s^2)c_m^+c_m$ with a
 finite width $s=2l_B$ to excite and trap either a quasihole or a
 quasiparticle. We consider a system of 20 electrons at $\nu = 2/3$,
 whose ground state angular momentum is $M = 280$ for $d=0.5l_B$. At
 $W_g = 0.2$ and $W_g = -0.2$, the global ground state resides in the
 angular momentum subspaces at $M=286$ and $M=274$, respectively. The
 change of the ground state angular momentum $M$ by $\pm 6$ suggests
 that we have induced a charge $|e|/3$ quasihole or a charge $-|e|/3$
 quasiparticle at the center of the 6-hole droplet.  The density plot
 for the electron ground state in Fig.~\ref{quasi} confirms the charge
 depletion and accumulation in the corresponding cases. The increase
 of the electron ground state angular momentum means a corresponding
 decrease in the angular momentum of the hole droplet, suggesting a
 quasiparticle excitation for the hole droplet and thus a quasihole
 excitation for the $\nu = 2/3$ electron ground state, as plotted in
 Figs.~\ref{quasi}(c) and (d). The decrease in angular momentum, on
 the other hand, suggests a quasiparticle excitation, as plotted in
 Figs.~\ref{quasi}(e) and (f). These excitations are localized at the
 origin, where we apply the impurity potential, and their presence has
 no effect on the edge excitation spectrum, since these are Abelian
 anyons. Roughly speaking, the quasihole and quasiparticle excitations
 induce the same density perturbation on the ground state, except for
 the opposite signs, suggesting a quasiparticle-quasihole symmetry.

 \section{concluding remarks}
 \label{sec:conclusion}

 To summarize, we study the ground states, edge, and bulk excitations
 of $\nu = 2/3$ fractional quantum Hall condensates in a
 semi-realistic microscopic model.  We find strong numerical evidence
 that a $\nu = 2/3$ droplet can be regarded as a $\nu_h = 1/3$
 Laughlin hole droplet embedded in a larger $\nu_I = 1$ integer
 quantum Hall droplet. In particular, we find two counter-propagating
 edge modes, which are associated with the inner edge (the edge of the
 hole droplet) and the outer edge (the edge of the integer quantum
 Hall edge). The inner edge mode is well separated from bulk
 excitations and resembles the edge of a $\nu = 1/3$ electron droplet,
 except that they propogate in opposite directions and respond
 oppositely to the edge confining potential, which explicitly breaks
 the particle- hole symmetry. The outer edge states have higher energy
 and mix with bulk excitations due to the computational limit on the
 Hilbert space dimension. The $\nu = 2/3$ quantum Hall droplet also
 has the same $\pm e/3$ quasiparticle and quasihole excitation as a
 $\nu = 1/3$ Laughlin droplet. These features are robust in the
 presence of finite electron layer thickness, which softens the
 Coulomb interaction between electrons.

 One of the major advantages of the disk geometry is that we can
 identify edge modes and determine the velocities of the edge
 modes. We have previously applied the same method to $\nu = 5/2$
 fractional quantum Hall systems and found significant differences in
 charge and neutral velocities,~\cite{wan:165316} which leads to quite
 different temperature regimes in which charge $e/4$ and charge $e/2$
 quasiparticles can be observed in interference
 experiments.~\cite{willett08,bishara:165302} Similarly, based on
 comparison with the edge excitations in the $\nu = 1/3$ Laughlin case
 and in the $\nu = 1$ integer case, we are able to resolve the edge
 excitations of the $\nu = 2/3$ case. It is interesting to point out
 that the outer edge velocity is roughly 3 times that of the inner
 edge mode velocity, which is about the same as the edge mode velocity
 of the $\nu = 1/3$ Laughlin liquid with the same Coulomb interaction
 and a similar confining potential strength due to the neutralizing
 charge background (Fig.~\ref{velocityd}c).  In general, edge mode
 velocities are non-universal, depending on details of
 electron-electron interaction and confining potential. For $\nu =
 2/3$, at which there are two counter-propagating edge modes, the
 velocities further depend on the coupling between the two edge modes
 affected by interactions and
 impurities,~\cite{wen:IJMPB,PhysRevLett.72.4129} for example. The
 fact that we are observing a relatively robust ratio ($\sim$3) of the
 outer edge mode velocity to the inner one (consistent with their
 density changes) suggests that, at the length scale of our
 finite-size study, the velocities are dominated by the Coulomb
 interaction strength determined by the electron density change
 associated with each mode and that the two edge modes are very weakly
 coupled. We note that, in the thermodynamic limit, the long-range
 nature of the Coulomb interaction is expected to force the two modes
 to reorganize into a charge mode and a neutral mode, with the charge
 mode velocity logarithmically divergent while the neutral mode
 velocity remains finite in the long wave length limit. The systems
 size of our study is too small to see this trend.

 Another feature of the current calculation is that we consider a
 semi-realistic confining potential arising from the neutralizing
 background charge. This, together with long-range interaction between
 electrons, allows us to compare the energetics of competing states
 and discuss the qualitative dependence of eigenenergies and edge
 mode velocities on the confining potential. This is an extremely
 interesting subject, especially when we have the $\nu = 5/2$ quantum
 Hall systems in mind. In that case, there are at least two competing
 candidates for the ground states, the Moore-Read Pfaffian state and
 its particle-hole conjugate, the anti-Pfaffian state. They are
 exactly degenerate, if we neglect the particle-hole symmetry
 breaking terms, such as the 3-body interaction due to Landau level
 mixing. As demonstrated in this study, the confining potential is
 also a relevant symmetry breaking term, which leads to an opposite dependence
 of the edge mode velocities on this potential. It has been
 predicted~\cite{wan:165316} that the anti-Pfaffian state is favored
 in weak confinement (smooth edge) while the Moore-Read Pfaffian
 state is favored in strong confinement (sharp edge). It is also
 interesting to point out that for $\nu = 2/3$ the dependence of the
 edge mode velocity of the inner edge on the confining potential is
 significantly weaker than that of the outer edge
 (Fig.~\ref{out_velocity}), suggesting a screening effect by the
 outer edge on the inner edge to the change of the confining potential.

 \section{acknowledgement}
 The authors are grateful to Matthew Fisher for discussions and
 encouragement that initiated this work, and to Kwon Park and Bernd
 Rosenow for helpful discussions. This work is supported by NSFC Grant
 No. 10504028 (Z.X.H., H.C. \& X.W.)  and NSF grants No. DMR-0704133
 (K.Y.) and DMR-0606566 (E.H.R.). This research was supported in part
by the PCSIRT (Project No. IRT0754). Z.X.H. thanks the Ministry of
 Education of China for support to visit the NHMFL. X.W. acknowledges
 the Max Planck Society (MPG) and the Korea Ministry of Education,
 Science and Technology (MEST) for the support of the Independent
 Junior Research Group at the Asia Pacific Center for Theoretical
 Physics (APCTP).

 \end{document}